\documentclass[conference]{IEEEtran}
\usepackage{cite}
\usepackage[bookmarks=true, bookmarksnumbered=true, bookmarksopen=true,
   unicode=true, colorlinks=true,
		pagebackref=true,
	linkcolor=blue,citecolor=blue,filecolor=blue,urlcolor=blue,
	pdfstartview=FitH
]{hyperref}
\usepackage[disable]
	{todonotes}

\usepackage{color}

\ifCLASSINFOpdf
\else
\fi
\usepackage{amsmath}
\usepackage{array}
\usepackage[caption=false,font=footnotesize]{subfig}
\usepackage{url}
\begin{document}

\title{Narrative Smoothing: Dynamic Conversational Network for the Analysis of TV Series Plots}

% \author{\IEEEauthorblockN{Michael Shell}
% \IEEEauthorblockA{School of Electrical and\\Computer Engineering\\
% Georgia Institute of Technology\\
% Atlanta, Georgia 30332--0250\\
% Email: http://www.michaelshell.org/contact.html}
% \and
% \IEEEauthorblockN{Homer Simpson}
% \IEEEauthorblockA{Twentieth Century Fox\\
% Email: homer@thesimpsons.com}
% \and
% \IEEEauthorblockN{James Kirk\\ and Montgomery Scott}
% \IEEEauthorblockA{Starfleet Academy\\
% San Francisco, California 96678--2391\\
% Telephone: (800) 555--1212\\


\author{\IEEEauthorblockN{Xavier Bost\IEEEauthorrefmark{1},
Vincent Labatut\IEEEauthorrefmark{1},
Serigne Gueye\IEEEauthorrefmark{1} and
Georges Linar\`es\IEEEauthorrefmark{1}}
\IEEEauthorblockA{\IEEEauthorrefmark{1}Laboratoire Informatique d'Avignon, EA 4128~--~University of Avignon, France\\ Email: \{firstname\}.\{lastname\}@univ-avignon.fr}}
%Telephone: (800) 555--1212, Fax: (888) 555--1212}


\maketitle

\begin{abstract}
  Modern popular \textsc{tv} series often develop complex storylines spanning several seasons, but are usually watched in quite a discontinuous way. As a result, the viewer generally needs a comprehensive summary of the previous season plot before the new one starts. The generation of such summaries requires first to identify and characterize the dynamics of the series subplots. One way of doing so is to study the underlying social network of interactions between the characters involved in the narrative. The standard tools used in the Social Networks Analysis field to extract such a network rely on an integration of time, either over the whole considered period, or as a sequence of several time-slices. However, they turn out to be inappropriate in the case of \textsc{tv} series, due to the fact the scenes showed onscreen alternatively focus on parallel storylines, and do not necessarily respect a traditional chronology. This makes existing extraction methods inefficient to describe the dynamics of relationships between characters, or to get a relevant instantaneous view of the current social state in the plot. This is especially true for characters shown as interacting with each other at some previous point in the plot but temporarily neglected by the narrative. In this article, we introduce \textit{narrative smoothing}, a novel, still exploratory, network extraction method. It smooths the relationship dynamics based on the plot properties, aiming at solving some of the limitations present in the standard approaches. In order to assess our method, we apply it to a new corpus of 3 popular \textsc{tv} series, and compare it to both standard approaches. Our results are promising, showing narrative smoothing leads to more relevant observations when it comes to the characterization of the protagonists and their relationships. It could be used as a basis for further modeling the intertwined storylines constituting \textsc{tv} series plots.
\end{abstract}


\vspace{2mm}

\noindent \textcolor{red}{\textbf{Cite as:}\\X.~Bost, V.~Labatut, S.~Gueye, G.~Linar\`es.\\\href{https://ieeexplore.ieee.org/document/7752379}{Narrative Smoothing: Dynamic Conversational Network for the Analysis of TV Series Plots.}\\DyNo: 2nd International Workshop on Dynamics in Networks, in conjunction with the 2016 IEEE/ACM International Conference ASONAM.\\doi: \href{https://doi.org/10.1109/ASONAM.2016.7752379}{10.1109/ASONAM.2016.7752379}}

% \fi
\IEEEpeerreviewmaketitle

%%%%%%%%%%%%%%%%%%%%%%%%%%%%%%%%%%%%%%%%%%%%%%%%%%%%%%%%%%%%%%%%%%%%%%%%
\section{Introduction}
% This demo file is intended to serve as a ``starter file''
% for IEEE conference papers produced under \LaTeX\ using
% IEEEtran.cls version 1.8b and later.
% I wish you the best of success.

 

\label{sec:intro}
\textsc{tv} series became increasingly popular these past ten years. As opposed to classical \textsc{tv} series containing standalone episodes with self-contained stories, modern series tend to develop continuous, possibly multiple, storylines spanning several seasons. However, the new season of a series is generally broadcast on a relatively short period: the typical dozen of episodes it contains is usually aired over a couple of months. In the most extreme case, the whole season is even released at once. Furthermore, modern technologies, like streaming or downloading services, tend to free the viewers from the broadcasting pace, often resulting in an even shorter viewing time (``binge-watching''). In summary, \textit{modern} \textsc{tv} series are highly continuous from a narrative point of view, but are usually watched in quite a discontinuous way: no sooner is the viewer hooked on the plot than he has to wait for almost one year before eventually knowing what comes next.

The main effect of this unavoidable waiting period is to make the viewer forget the plot, especially when complex. Since he fails to remember the major events of the previous season, he needs a comprehensive recap before being able to fully appreciate the new season. Such recaps come in various flavors: textual synopsis of the plot sometimes illustrated by keyframes extracted from the video stream; extractive video summaries of the previous season content, such as the ``official'' recap usually introduced at the beginning of the very first episode of the new season; or even videos of fans reminding, when not commenting, the major narrative events of the previous season. Though quite informative and sometimes enjoyable, such content-oriented summaries of complex plots always rely on a careful human expertise, usually time-consuming. The question is therefore to know how this task can be partially or even fully automated.

To the best of our knowledge, few works in the multimedia processing field focused on automatically modeling the plot of a movie. In~\cite{Guha2015}, the authors make use of low-level, stylistic features in order to automatically detect the typical three-act narrative structure of Hollywood full-length movies. Nonetheless, such a style-based approach does not provide any insight into the story content and focuses on a fixed narrative structure that generalizes with difficulty to the complex plots of \textit{modern} \textsc{tv} series. The benefits of Social Networks Analysis (\textsc{sna}) for investigating the plot content of fictional works have recently been emphasized in several articles. Most focus on literary works: dramas~\cite{Moretti2011a}, novels~\cite{Agarwal2012}, etc. In the context of multimedia works, \textsc{sna}-based approaches are even more recent and sparser \cite{Weng2007, Weng2009, Ercolessi2012}. However, these works focus either on full length-movies or on standalone episodes of classical \textsc{tv} series, where character interactions are often well-structured into stable communities. These approaches consequently do not necessarily translate well when applied to \textit{modern} \textsc{tv} series.

In this paper, we present an \textsc{sna}-based method aiming at automatically providing some insight into the complex plots of \textsc{tv} series, while solving the limitations of the previous works. For this purpose, we do consider not only standalone episodes or full-length movies with stable and well-defined communities, but the complex plots of \textsc{tv} series, as they evolve over dozens of episodes. In this case, no prior assumption can be made about a stable, static community structure that would remain unchanged in every episode and that the story would only uncover, and we have to deal with evolving relationships, possibly temporarily linked into dynamic communities. In this case, we are left with building the current state of the relationships upon the story itself, which, by focusing alternatively on different characters in successive scenes, prevents us from monitoring instantaneously the full social network underlying the plot. We thus propose to address this problem by smoothing the sequentiality of the narrative, resulting in an instantaneous monitoring of the current state of any relation at some point of the story. Our main contributions are the following. The first is \textit{narrative smoothing}, the method we propose for the extraction of dynamic social networks of characters. The second is the annotation of a corpus of $96$ \textsc{tv} series episodes from three popular \textsc{tv} shows: \textit{Breaking Bad}, \textit{Game of Thrones}, and \textit{House of Cards}. The third is a preliminary evaluation of our framework on these data, and a comparison with existing methods.

The rest of the article is organized as follows. In Section~\ref{sec:review}, we review in further details the previous works related to \textsc{sna}-based plot identification. Then, in Section~\ref{sec:methods}, we describe the method we propose. We first focus on the way the verbal interactions between characters are estimated, before detailing the way a dynamic view of the relationships in the plot of a \textsc{tv} series can be built independently from the narrative pace. In Section~\ref{sec:exp}, we illustrate how our tool can be used by applying it to the three mentioned \textsc{tv} series, and we compare the obtained results to existing methods.

%%%%%%%%%%%%%%%%%%%%%%%%%%%%%%%%%%%%%%%%%%%%%%%%%%%%%%%%%%%%%%%%%%%%%%%%
\section{Previous Works}
\label{sec:review}
In our review, we distinguish works considering a static network resulting from the temporal integration over the whole considered period, which we call \textit{complete aggregation}, from those extracting and studying a dynamic network based on a sequence of smaller integration periods called \textit{time-slices}.

%%%%%%%%%%%%%%%%%%%%%%%%%%%%%%%%%%%%%%%%
\subsection{Complete Aggregation}
\label{sec:cum_net}
Cumulative networks were widely used when attempting to apply \textsc{sna} for analyzing the plot of fictional works. The interactions are iteratively inserted as edges in the network of characters. They are possibly weighted and even directed, resulting in a static graph agglomerating every past relationship, whatever their time ordering.

In~\cite{Moretti2011a}, Moretti underlines and illustrates the light \textsc{sna} can shed on literary works, either plays or novels. By projecting the time of the character interactions onto the plane of a graph, \textsc{sna} helps to unveil some underlying patterns invisible to a closer reading. Moretti agglomerates the conversational interactions between characters in Shakespeare's \textit{Hamlet}, and exhibit the contrast between the Court, densely connected, and the emerging modern State, weakly connected around Horatio.

In~\cite{Weng2007, Weng2009}, relying on similar observations, Weng \textit{et al}. make use of \textsc{sna} to automatically analyze the plot of a movie. The social network of characters (denoted ``RoleNet'') is built as follows. They first manually characterize the scenes by their boundaries and the characters they involve. They then hypothesize an interaction between two characters whenever they both appear within the same scene. The network is obtained by representing characters as nodes and their interactions by links. These links are weighted according to the number of scenes in which they co-appear, resulting in a \textit{cumulative} representation of time. The authors analyze this network through community detection. They apply this approach to so-called ``bilateral movies'', which involve only two major characters, each of them central in his own community. In~\cite{Weng2007}, the \textit{RoleNet} is used for further investigating the plot, by classifying scenes into one of the two storylines constituting a bilateral movie. In~\cite{Weng2009}, an extended version of the network, without any prior assumption about the number of communities involved, is used as a basis for automatically detecting breakpoints in the story: a narrative breakpoint is assumed if the characters involved in successive scenes are socially distant in the network of characters accumulated over the whole story.

In~\cite{Ercolessi2012}, a similar network of interacting speakers is used, among other features, for clustering scenes of two \textsc{tv} series episodes into separate storylines, defined as homogeneous narrative sequences related to major characters. A standard community detection algorithm is applied to the network of speakers, as built upon each episode, before the social similarity between any pair of scenes is computed, as a relevant high-level feature for clustering scenes into sub-stories.

In summary, cumulative networks can be used as a reliable basis for automatically or manually analyzing the plot of fictional works with well-defined communities, as in dramas, full-length movies or standalone episodes of classical \textsc{tv} series\footnote{The website \cite{Kaminski2012} designed by Kaminski \textit{et al.} provides a convenient way of interactively visualizing such cumulative character networks for a database of about 700 movies.}.
But for \textsc{tv} series with complex, evolving and possibly parallel storylines, such a static approach is not appropriate. Indeed, a cumulative network built over a long period of time, as in modern \textsc{tv} series, is relatively dense and does not enable to extract meaningful information. Moreover, communities in the final agglomerative network undoubtedly always correspond to sub-stories, partially disconnected in the narrative, but the opposite does not generally stand. Some individuals may have been strongly connected to each other at some point of the story, before some of them interact with other people for some time, resulting in a second sub-story. Once agglomerated in the cumulative network, such changes in the interaction patterns may be obscured. In some extreme cases, distinct narrative sequences may even result in a complete cumulative graph, for instance in the interaction pattern that follows: $s_{12}^{(1)}...s_{12}^{(2)}s_{13}^{(3)}...s_{13}^{(4)}s_{23}^{(5)}...s_{23}^{(6)}$, where $s_{ij}^{(t)}$ denotes the fact that the $i$\textsuperscript{th} and $j$\textsuperscript{th} characters are the only interacting speakers in the $t$\textsuperscript{th} episode. The three consecutive interaction sequences result in a triangular interaction pattern unable to reflect the three corresponding sub-stories.

%%%%%%%%%%%%%%%%%%%%%%%%%%%%%%%%%%%%%%%%
\subsection{Time-slices}
\label{subsec:time_slice}
Some works attempt to take into account the evolution of the social network of the characters when analyzing the plot of fictional works. In~\cite{Agarwal2012}, the authors emphasize the limitations of the static, cumulative graph when analyzing the centrality of the various characters of the novel \textit{Alice in Wonderland}. A dynamic view of the social network is then introduced, by building successive static views of the network in every chapter, before standard centrality measures are separately computed in each of them and traced over time for some major characters. Each view corresponds to a so-called \textit{time-slice}.

Though widely used \cite{Holme2012} when considering the evolution over time of general networks (\textit{i.e.} not necessarily narrative ones), time-slice networks, as resulting from the differentiation over some time step of the cumulative network, may still be problematic. In~\cite{Clauset2012}, the authors focus on the critical issue of the time-slice duration, called ``snapshot rate''. It must be chosen carefully to allow to capture a sufficient amount of interactions, but not too many, otherwise one may obtain irrelevant network statistics. The authors then describe a way of automatically estimating the natural time-slice for monitoring over time the evolution of a network of daily contacts in a professional context.

In order to monitor the plot of \textsc{tv} series and allow further analysis, such a time-slice should be short enough to capture punctual narrative events related to the social network of characters, but long enough to provide a comprehensive view of the relationships state at any point of the story. Unfortunately, getting such a snapshot of the current state of the relationships between the protagonists within the plot turns out to be particularly challenging.

As a smoother alternative for monitoring the state of the network over time, \cite{Mutton2004} applies temporal decay to the past occurrences of the relationships between the characters of Shakespeare's plays for monitoring their evolution over time.

\begin{figure}[!t]
  \centering
  \includegraphics[width=.49\textwidth]{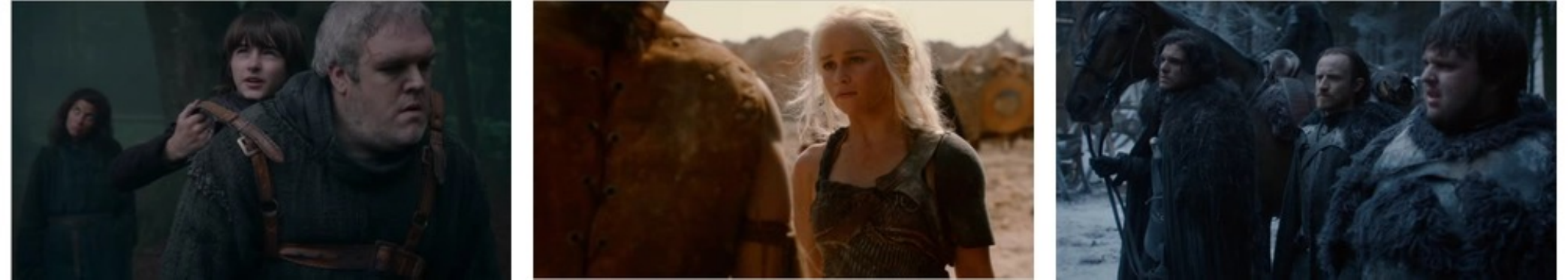}
  \caption{\label{fig:first_seq}Three different sets of interacting characters from three consecutive scenes.}
\end{figure}

Unlike the network of physical contacts described in~\cite{Clauset2012}, the state of the relationships within a story is not fully monitored at any moment, but has to be inferred from the story itself. The narrative usually focuses alternatively on some relationships, possibly belonging to parallel storylines, and only provides a partial view on the network's current state. Some relationships may even take place at the same moment in different places, but will be shown sequentially in successive scenes. Fig.~\ref{fig:first_seq} illustrates the typical sequential nature of the story as being narrated: three disjoint sets of interacting speakers, possibly at the same time but in different places, are shown sequentially in the story in three successive scenes.

\begin{figure}[!t]
  \centering
  \includegraphics[width=.45\textwidth]{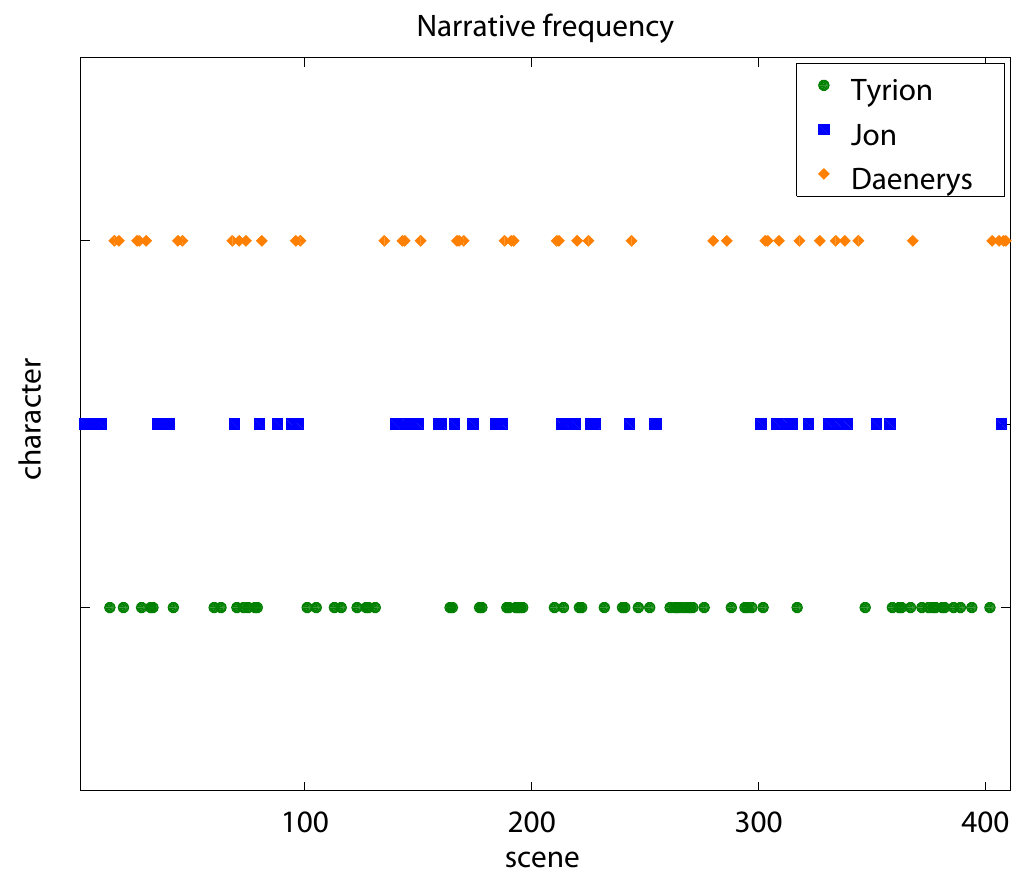}
  \caption{\label{fig:narr_freq}Narrative frequency of three character-based storylines in the first two seasons of \textit{Game of Thrones}.}
\end{figure}

As a consequence, the temporalness of the narrative may be quite different from the temporalness of the underlying network: in particular, the mere fact that a group of mutually interacting characters temporarily disappears from the story does not imply that the corresponding relationships disappeared from the network. The narrative focus on those relationships may only have been postponed by the filmmaker. Furthermore, the pace of activation of the relationships occurring in different regions of the interaction network remains largely unpredictable, especially when multiple, disjoint storylines take place in parallel within the narrative. Fig.~\ref{fig:narr_freq} plots the scene occurrences of $3$ major character-based storylines in the first two seasons of \textit{Game of Thrones}. Except in the very beginning of the first season, where Jon and Tyrion meet each other, the $3$ characters interact within well-separated communities.

As can be seen, the way the story alternatively activates these three major storylines does not seem to follow some regular patterns. In such a case, the ``ideal'' time-slice may be tricky to set. If too large, it will possibly mask the fast changes usually occurring in the most frequently activated storyline, here the story centered around Tyrion. If too narrow, it would lead to irrelevant interpretations of the narrative disappearance of some groups of relationships: the absence of Jon Snow's storyline from the scene 250 up to the scene 300 does definitely not imply that he does not remain socially active in the meantime in his own community. Therefore, the sequential nature of the story should prevent us to identify the time of the narrative to the time objectively affecting the social network that the story sequentially unveils.

In the rest of this paper, we introduce a novel way of building the dynamic network of interactions between the characters of \textsc{tv} series that allows to fully capture the instantaneous state of every relationship at any point of the story, whatever the pace of activation of each storyline in the narrative.

%%%%%%%%%%%%%%%%%%%%%%%%%%%%%%%%%%%%%%%%%%%%%%%%%%%%%%%%%%%%%%%%%%%%%%%%
\section{Methods}
\label{sec:methods}
We now describe the two steps constituting our method\footnote{Source code available online at: \href{https://github.com/bostxavier/Narrative-Smoothing}{github.com/bostxavier/Narrative-Smoothing}}. First, we explain how we identify and characterize interactions between \textsc{tv} series characters. Second, we describe how we extract a smoothed dynamic network from a set of interactions.

%%%%%%%%%%%%%%%%%%%%%%%%%%%%%%
\subsection{Estimating Verbal Interactions}
\label{subsec:interactions}
In this work, we focus on relationships defined in a \textit{strong sense}, as based on personal, verbal, interactions between characters. The resulting network can thus be considered as a \textit{conversational network}, in contrast to the co-occurrence network of characters described in~\cite{Weng2007,Weng2009} and used in~\cite{Ercolessi2012}.

\begin{figure}[!t]
  \centering
  \includegraphics[width=.48\textwidth]{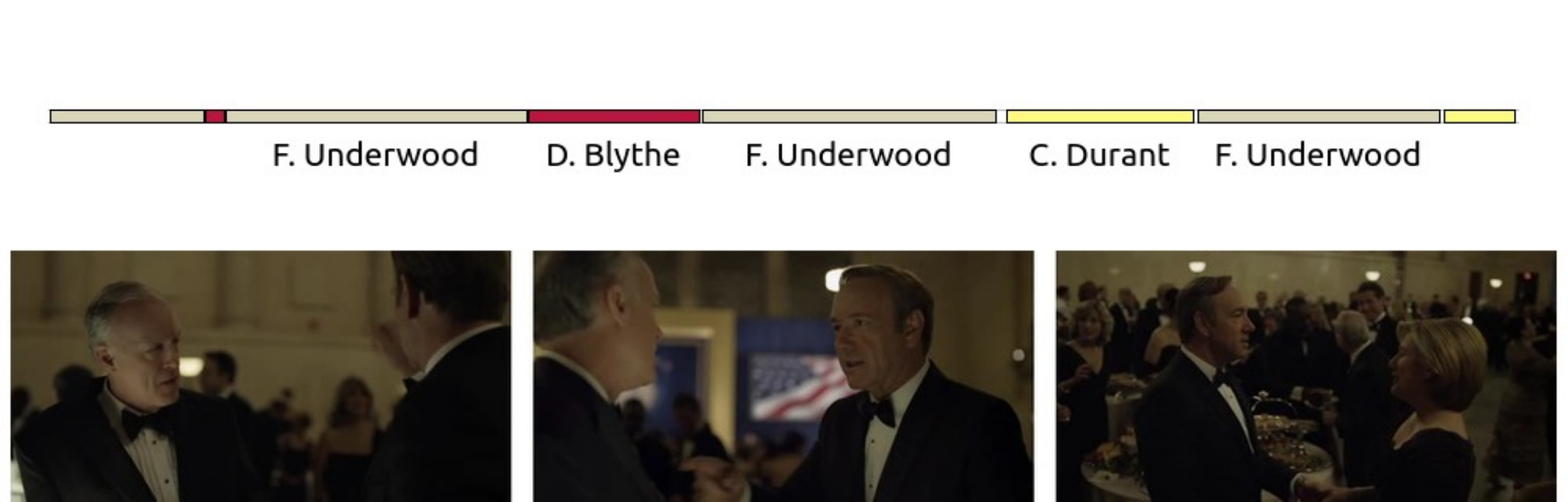}
  \caption{\label{fig:irrelSeq}Two consecutive dialogue sequences
    within the same scene.}
\end{figure}

We first manually annotated the scenes boundaries: similarly to the rule of the three unities classically prescribed for dramas, a scene in a movie is defined as a homogeneous sequence of actions occurring at the same place, within a continuous period of time. The characters co-appearing in a single scene are thus supposed to interact with one another. However, if being at the same place at the same time is usually required to consider that several persons interact, it is rarely sufficient. Fig.~\ref{fig:irrelSeq} shows two consecutive dialogues extracted from the \textsc{tv} series \textit{House of Cards}, and belonging to the same scene. Three speakers are involved, but without any interaction between the second (\textit{D.~Blythe}) and the third (\textit{C.~Durant}) ones. The first speaker (\textit{F.~Underwood}) is talking to \textit{D.~Blythe} in the first sequence, then is moving to \textit{C.~Durant} and starts discussing with her.

\begin{figure}[!t]
  \begin{tabular}{cc}
  \subfloat[Rule~\textit{(1)}] {
    \includegraphics[width=.22\textwidth]{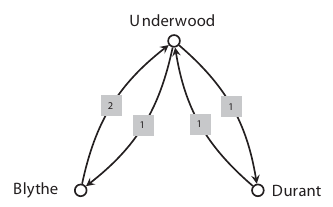}
    \label{subfig:rule_1}
  }
  &
  \subfloat[Rule~\textit{(2)}] {
    \includegraphics[width=.22\textwidth]{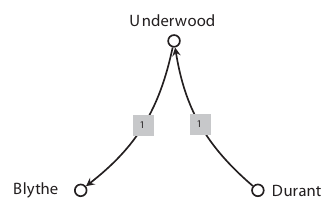}
    \label{subfig:rule_2}
  }
  \\
  \subfloat[Rule~\textit{(4)}] {
    \includegraphics[width=.22\textwidth]{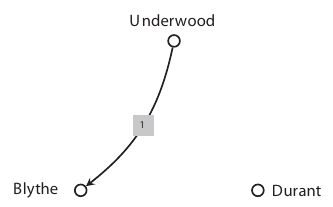}
    \label{subfig:rule_4}
  }
  &
  \subfloat[Rules~\textit{(1--4)}] {
    \includegraphics[width=.22\textwidth]{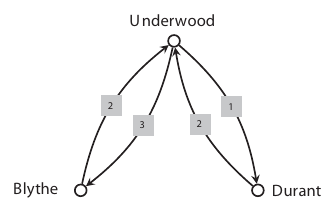}
    \label{subfig:all_rules}
  }
  \\
  \end{tabular}
  \caption{\label{fig:net_rules}Verbal interactions estimated from the
    separate and joint application of Rules~\textit{(1, 2, 4)} to the speech
    turn sequence shown on Fig.~\ref{fig:irrelSeq}.}
\end{figure}

Instead of globally considering the scene unit, we chose to tackle this problem by identifying the verbal interactions upon the sequence of speech turns in each scene, once manually labeled according to the corresponding speakers. In order to estimate the verbal interactions from the single sequence of utterances, we apply four basic heuristics:

\vspace{2mm}
\noindent \textbf{Rule~\textit{(1)}: Surrounded speech turn.} We consider that a speaker $s_2$ is talking to another speaker $s_1$ if he is speaking both after and before him, resulting in a speech turns sequence $s_1s_2s_1$, where each speech turn is labeled according to the corresponding speaker. Fig.~\ref{subfig:rule_1} shows the subgraph resulting from the application of Rule~\textit{(1)} to the speech turns sequence shown on Fig.~\ref{fig:irrelSeq}, where each edge is labeled according to the number of times each speaker is considered as talking to another one.

\vspace{2mm}
\noindent \textbf{Rule~\textit{(2)}: Starting and ending speech turns.} This rule aims at processing the first and last utterances of each sequence $s_1s_2...s_3s_4$ of speech turns, by adding two links $s_1 \rightarrow s_2$ from the first to the second speaker and $s_4 \rightarrow s_3$ from the fourth to the third one. The network resulting from the application of Rule~\textit{(2)} to the sequence of Fig.~\ref{fig:irrelSeq} is shown on Fig.~\ref{subfig:rule_2}.

The last two rules are introduced to process ambiguous sequences of the type $s_1s_2s_3$, where three consecutive speech turns originate in three different speakers: in such cases, the second speaker could be stated as talking to the first one as well as to the third one, or even to both of them. However, such speech turns sequences can often be disambiguated by considering speakers preceding and following the sequence.

\vspace{2mm}
\noindent \textbf{Rule~\textit{(3)}: Local disambiguation.} We distinguish 2 variants of this rule. Rule~\textit{(3a)} applies when the second speaker appears before the sequence, but not after, as in $(s_2)s_1s_2s_3(s_4)$. We then consider that $s_2$ is speaking with $s_1$ rather than with $s_3$. Symmetrically, Rule~\textit{(3b)} concerns the case when the second speaker appears after, but not before the sequence, as in $(s_0)s_1s_2s_3(s_2)$, and is therefore assumed to speak to $s_3$.

\vspace{2mm}
\noindent \textbf{Rule~\textit{(4)}: Temporal proximity.} When the second speaker is involved in the conversation both before and after the ambiguous sequence, as in $(s_2)s_1s_2s_3(s_2)$, we consider the ambiguous speech turn to be intended for the speaker whose utterance is temporally closer. In the sequence shown on Fig.~\ref{fig:irrelSeq}, the fifth, ambiguous utterance would then be hypothesized as intended for the first speaker \textit{D.~Blythe}, resulting in the additional link shown on Fig.~\ref{subfig:rule_4}. The same Rule~\textit{(4)} is applied when the speaker $s_2$ is not involved in the immediate conversational context.

Fig.~\ref{subfig:all_rules} shows the amount of directed interactions between any two speakers involved in the scene shown on Fig.~\ref{fig:irrelSeq}. For the sake of simplicity, the interactions are evaluated here as a number of interactions from one speaker to another one, but can as well be expressed in terms of interaction duration.

%%%%%%%%%%%%%%%%%%%%%%%%%%%%%%
\subsection{Extracting the Dynamic Network}
\label{subsec:net_construct}
We obtain the total amount of interaction $h_{ij}^{(t)}$ between the speakers $s_i$ and $s_j$ in the $t$\textsuperscript{th} scene by summing up the amount of speech flowing in the scene $t$ from $s_i$ to $s_j$ and from $s_j$ to $s_i$, resulting in an undirected local interaction amount, possibly null, expressed in seconds.

As stated in Section~\ref{sec:intro}, we would like to get an instantaneous measurement of the strength of any relationship at any moment, but from the successive partial views of the underlying network that the narrative provides us. Intuitively, a particular relationship may be considered as especially important at some point of the story if the involved characters both speak frequently and a lot to each other: the time interval needed before the interaction is reactivated in the narrative is expected to be short, and the interaction time to be long whenever the relationship is active in the plot.

Four possible states have to be considered when monitoring a single relationship over time: (1) the relationship is active in the current scene; (2) it has been active in the story and will be active again later; (3) it was active before, but will no longer be active in the narrative; and (4) it has not yet been active in the narrative.

The first case is the simplest one: each time the interaction occurs, its strength can be estimated in a standard way as the duration of the interaction, expressed in seconds: at any scene $t$ where speakers $s_i$ and $s_j$ are hypothesized as talking to each other, the instantaneous weight of their relationship $w_{ij}^{(t)}$ is estimated as follows:
\begin{equation}
  w_{ij}^{(t)} = h_{ij}^{(t)}
  \label{eq:inter}
\end{equation}

\noindent where $h_{ij}^{(t)}$ denotes the interaction time between the $i$\textsuperscript{th} and $j$\textsuperscript{th} speakers in scene $t$.

The last three cases are much trickier. Between two consecutive occurrences of the same relationship in the story, it would be tempting to consider that the relationship is still (resp. already) active if it is recent (resp. imminent) enough at each moment considered. In~\cite{Mutton2004}, the author applies temporal decay to the past occurrences of the interactions between characters in Shakespeare's plays in order to visualize their evolution over time. According to the time-slice framework described in Section~\ref{sec:review}, as long as the relationship is present in the observation window of the network over time, it is stated as active, and inactive as soon as no longer observed.

As emphasized in Section~\ref{sec:review}, such a way of handling the past and future occurrences of the relationships is inappropriate for many modern \textsc{tv} series. Some interacting characters may be absent of the narrative for an undefined period of time but still linked in the underlying network, as confirmed by the fact that the last state of the relationship is generally used as a starting point when the characters are re-introduced in the story. Indeed, the temporalness of the narrative should affect a relationship only when at least one of the involved characters interacts with others after and/or before the relationship is active: the relationship between two characters should only get weaker if they interact separately with others before interacting again with one another.

In order to perform such a narrative smoothing, we introduce two quantities to handle the scenes where the two characters do not interact. First, $\Delta_{ij}^{(l)}(t)$ is the \textit{narrative persistence} between speakers $s_i$ and $s_j$, considered at scene $t$. It is defined relatively to the last scene in which their relation was active, noted $l$:
\begin{equation}
  \Delta_{ij}^{(l)}(t) = h_{ij}^{(l)} - \sum_{t' = l + 1}^t  \sum_{k \neq i,j} \left ( h_{ik}^{(t')} + h_{jk}^{(t')} \right )
  \label{eq:narr_persist}
\end{equation}
This measure $\Delta_{ij}^{(l)}(t)$ corresponds to the net balance between the duration of the last interaction occurrence $h_{ij}^{(l)}$ and the conversational time (represented by the double sum) the two characters $i$ and $j$ have devoted separately to other characters $k$ since then.

Symmetrically, $\Delta_{ij}^{(n)}(t)$ is the \textit{narrative anticipation} between speakers $s_i$ and $s_j$, considered at scene $t$. It is defined relatively to the next scene in which their relation will be active again, noted $n$:
\begin{equation}
\Delta_{ij}^{(n)}(t) = h_{ij}^{(n)} - \sum_{t' = t}^{n-1} \sum_{k \neq i,j} \left ( h_{ik}^{(t')} + h_{jk}^{(t')} \right )
\label{eq:narr_antic}
\end{equation}

We then define the instantaneous weight $w_{ij}^{(t)}$ of the relationship between the speakers $s_i$ and $s_j$ in any scene $t$ occurring between two consecutive occurrences of their relationship as: 
\begin{equation} 
  w_{ij}^{(t)} = \max \left \{ \Delta_{ij}^{(l)}(t), \Delta_{ij}^{(n)}(t) \right \}
\label{eq:between}
\end{equation}
\noindent If neither of the two characters speaks to others before they interact again with one another, $w_{ij}^{(t)} = \max \left \{ h_{ij}^{(l)}, h_{ij}^{(n)} \right \}$ and the last (resp. next) occurrence of the relation is considered as still (resp. already) fully present in the network, whatever the number of intermediate scenes the narrative introduces in-between to focus on other parts of the plot.

The weight of the relationship between the $i$\textsuperscript{th} and $j$\textsuperscript{th} speakers in any scene $t$ occurring after its very last occurrence in the narrative is expressed as follows, provided that one of the two characters remains involved in the story by interacting with others:
\begin{equation} 
  w_{ij}^{(t)} = \Delta_{ij}^{(l)}(t)
\end{equation}

\noindent Symmetrically, the weight of the relationship between the $i$\textsuperscript{th} and $j$\textsuperscript{th} speakers in any scene $t$ occurring before its first occurrence in the story is computed as follows, as long as one the two characters has already been shown as interacting with other people:
\begin{equation} 
  w_{ij}^{(t)} = \Delta_{ij}^{(n)}(t)
\end{equation}

\noindent In the very last case, when neither of the two characters is still (resp. already) active, the weight $w_{ij}^{(t)}$ is set to $-\infty$.

We then normalize the weights of the interactions linking any couple of characters in some scene $t$. We use the following formula, resulting in an undirected graph $\mathcal{G}^{(t)}$, capturing the instantaneous state of the social network that the story sequentially unveils:
\begin{equation} 
  n_{ij}^{(t)} = \frac{1}{1 + e^{-\lambda w_{ij}^{(t)}}}
\end{equation}
where \noindent $n_{ij}^{(t)}$ is the normalized weight of the relationship between the speakers $s_i$ and $s_j$.

The choice of the sigmoid function for such a normalization purpose both allows to get weights ranging from 0 to 1 and to simulate the way the past and future states of a relationship in the narrative could influence its current state at some point $t$. The parameter $\lambda$ is a  parameter of sensitivity to the past and future states of the network and was set to $\lambda = 0.01$ (high values imply low dependence on the future and past states).

\begin{figure}[!t]
  \centering
  \includegraphics[width=.495\textwidth]{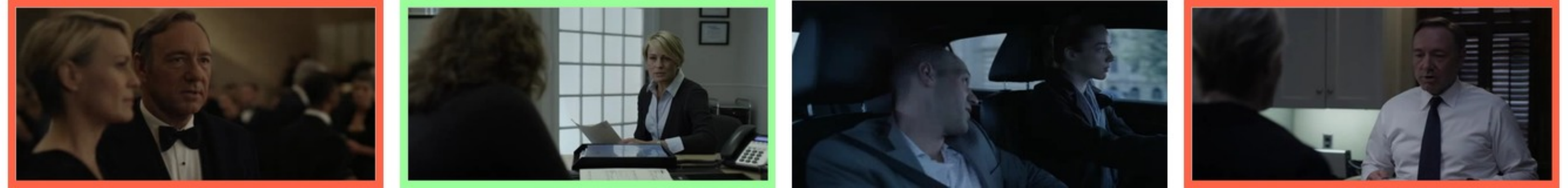}
  \caption{\label{fig:weighting_scheme}Example of application of the weighting scheme to a specific relationship.}
\end{figure}

Fig.~\ref{fig:weighting_scheme} shows four excerpts of four consecutive scenes in~\textit{House of Cards}, involving five individuals. The first two of them, namely \textit{Francis Underwood} and his wife \textit{Claire}, interact with each other in the first and last scenes (red border) respectively during 30 and 20 seconds, whereas Claire interacts in-between 40 seconds with another person in the second scene (green border) and two other people are talking to one another in the third scene during 50 seconds.

In the first and fourth scenes, Claire and Francis are interacting with each other: according to Equation~\ref{eq:inter}, we then set the weights of their relationship to the corresponding interaction times, respectively $30$ and $20$ seconds. In the second scene, the last interaction between Claire and Francis is on the one hand weakened by the separate interaction of Claire with someone else during $40$ seconds: the resulting \textit{narrative persistence} of the relationship between Francis and Claire then amounts to $\Delta_{12}^{(1)}(2) = 30 - 40 = -10$ (Equation~\ref{eq:narr_persist}). On the other hand, the \textit{narrative anticipation} with respect to the next interaction between Francis and Claire then amounts to  $\Delta_{12}^{(4)}(2) = 20 - 40 = -20$ (Equation~\ref{eq:narr_antic}), resulting in an instantaneous weight $w_{12}^{(2)} = \max \{-10, -20\} = -10$ in the second scene. In the third scene, neither of the two characters is involved: the narrative persistence of their relationship is unchanged, but the narrative anticipation then increases to 20, because no interfering character separates at this point Francis and Claire from their next interaction in the fourth scene. We then have $w_{12}^{(3)} = \max \{-10, 20\} = 20$ and the full resulting sequence of unnormalized, instantaneous weights for the relationship between Claire and Francis is then (30, -10, 20, 20) at the four considered moments.

%%%%%%%%%%%%%%%%%%%%%%%%%%%%%%%%%%%%%%%%%%%%%%%%%%%%%%%%%%%%%%%%%%%%%%%%
\section{Experiments and Results}
\label{sec:exp}
In this section, we qualitatively evaluate \textit{narrative smoothing}, our graph extraction method, by comparing it to both types of methods described in Section~\ref{sec:review}. For this purpose, we focus on three recent and popular \textsc{tv} series, and explore their plots from the dynamics of their underlying social network of characters. We first describe our corpus and then analyze the obtained networks from the perspective of the protagonists (nodes) and their relationships (links).

%%%%%%%%%%%%%%%%%%%%%%%%%%%%%%%%%%%%%%%
\subsection{Corpus}

\begin{table}[ht]
	\caption{\label{tab:corpus} Main features of each \textsc{tv} series: Breaking Bad (BB), Game of Thrones (GoT) and House of Cards (HoC).}
	\centering
	\begin{tabular}{| p{4.7cm} || r | r | r |}
    \hline
    Corpus & \textsc{BB} & \textsc{GoT} & \textsc{HoC} \\
    \hline
    \hline
    \# episodes & 20 & 50 & 26 \\
    Total duration (hours) & $\simeq$ 15 & $\simeq$ 42 & $\simeq$ 19 \\
     Speech duration (seconds) & 23,403 & 67,578 & 39,175 \\
    \# subtitles & 11,544 & 33,834 & 21,005 \\
    \hline
    \hline
    \# scenes & 402 & 1,073 & 912 \\
    \% spoken scenes & 95.03 & 96.36 & 97.70 \\
    \hline
    \hline
    \# speakers/scene (avg.) & 2.38 & 2.93 & 2.49 \\
    \# speakers/scene (std. dev.) & 1.16 & 1.60 & 1.12 \\
    \hline
  \end{tabular}
\end{table}

Our corpus consists in three very popular \textsc{tv} series: \textit{Breaking Bad} (first 2 seasons), \textit{Game of Thrones} (first 5 seasons), and \textit{House of Cards} (first 2 seasons). We manually annotated the scene boundaries and labeled each subtitle according to the corresponding speaker. The obtained annotations were then used to extract the social networks of characters, by first estimating the verbal interactions according to the rules described in Subsection~\ref{subsec:interactions} and then by using the existing methods presented in Section~\ref{sec:review} as well as our own narrative smoothing approach. The resulting networks are publicly available online\footnote{\url{https://dx.doi.org/10.6084/m9.figshare.2199646}}, along with short videos showing the evolution of the three networks of characters over the seasons considered. Table~\ref{tab:corpus} reports the main features of the resulting corpus.

Speech is uniformly distributed over the scenes, with in average more than $95\%$ of the scenes containing at least one subtitle, which suggests that most social interactions are expressed verbally in these three \textsc{tv} series. Furthermore, the average number of speakers by scene remains quite low (ranging from $2.38$ to $2.93$ depending on the \textsc{tv} series), often resulting in the simple patterns of verbal interactions properly handled by applying the basic heuristics described in Subsection~\ref{subsec:interactions}.

%%%%%%%%%%%%%%%%%%%%%%%%%%%%%%%%%%%%%%%
\subsection{The protagonists}
We first base our analysis on the protagonists of the considered \textsc{tv} series, \textit{i.e.} the nodes in the corresponding extracted social networks. We present only a small number of results, which concern characters of particular interest. We characterize them using the \textit{node outgoing strength}, a generalization of the node degree defined as the sum of the weights of the links originating from the considered node. In our case, weights correspond to spoken interaction durations, either normalized (narrative smoothing) or not (cumulative and time-slice based approaches): the strength of a character is thus related to how much he speaks to others.

\begin{figure}[!t]
  \centering
  \includegraphics[width=.43\textwidth]{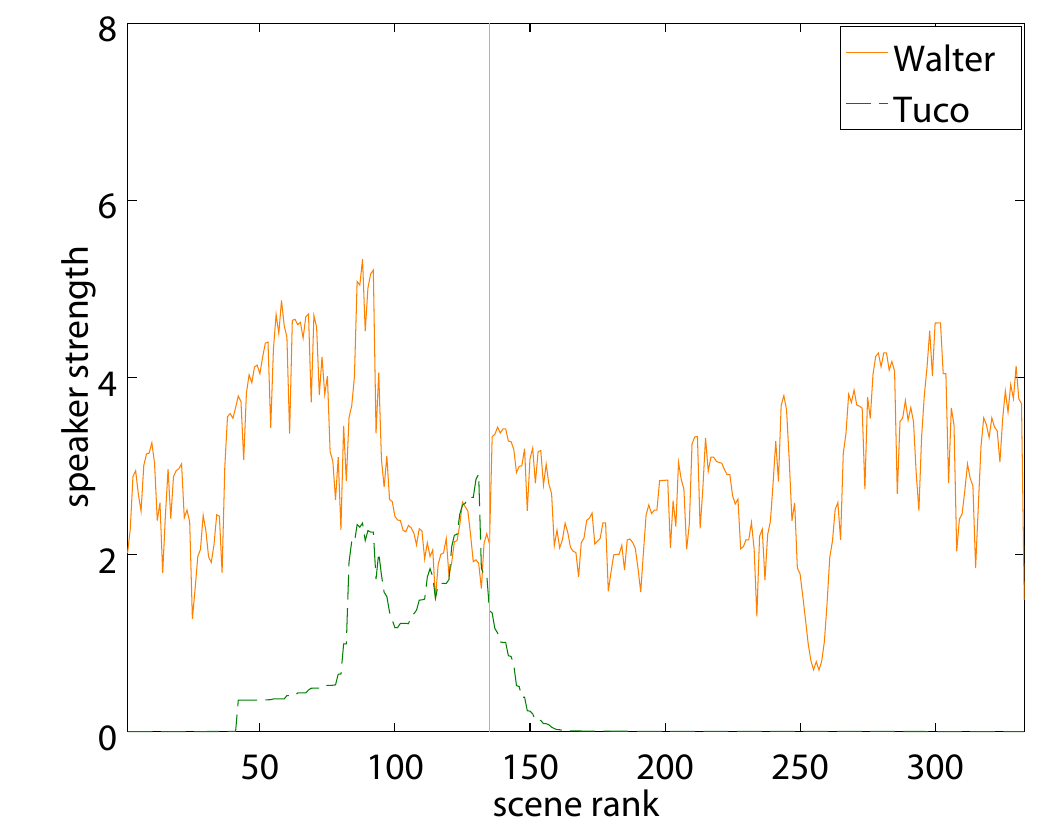}
  \caption{\label{fig:tuco_walt_dyn}Strengths of two important characters in~\textit{Breaking Bad}, plotted as a function of the scenes.}
\end{figure}

We first focus on Walter White, the main character of \textit{Breaking Bad}, and Tuco Salamanca, one of the drug dealers with whom he is in business. The cumulative network (as described in Section \ref{sec:cum_net}), \textit{i.e.} the temporal integration over the first $20$ episodes, is represented in Fig.~\ref{fig:BB_cumul} (Appendix). In this network, the strength of Walter White (his total interaction time with others) is about twenty times as large as the strength of Tuco: $12,332$ seconds for Walter (rank~1) \textit{vs.} $590$ for Tuco (rank~11). By comparison, Fig.~\ref{fig:tuco_walt_dyn} displays the evolution of their instantaneous strengths, obtained with our narrative smoothing method, as a function of the scenes ordered chronologically. This leads us to a completely different vision of Tuco's role in the plot. As Fig.~\ref{fig:tuco_walt_dyn} shows, from scene 100, his importance tends to increase and even overcomes the importance of the main protagonist for some time, before suddenly decreasing after scene 130. This clearly corresponds to a subplot, or a short narrative episode, ending with Tuco's death, at the end of scene 135 (vertical line on Fig.~\ref{fig:tuco_walt_dyn}).

\begin{figure}[!t]
  \centering
  \begin{tabular}{c}
  \subfloat[] {
    \includegraphics[width=.43\textwidth]{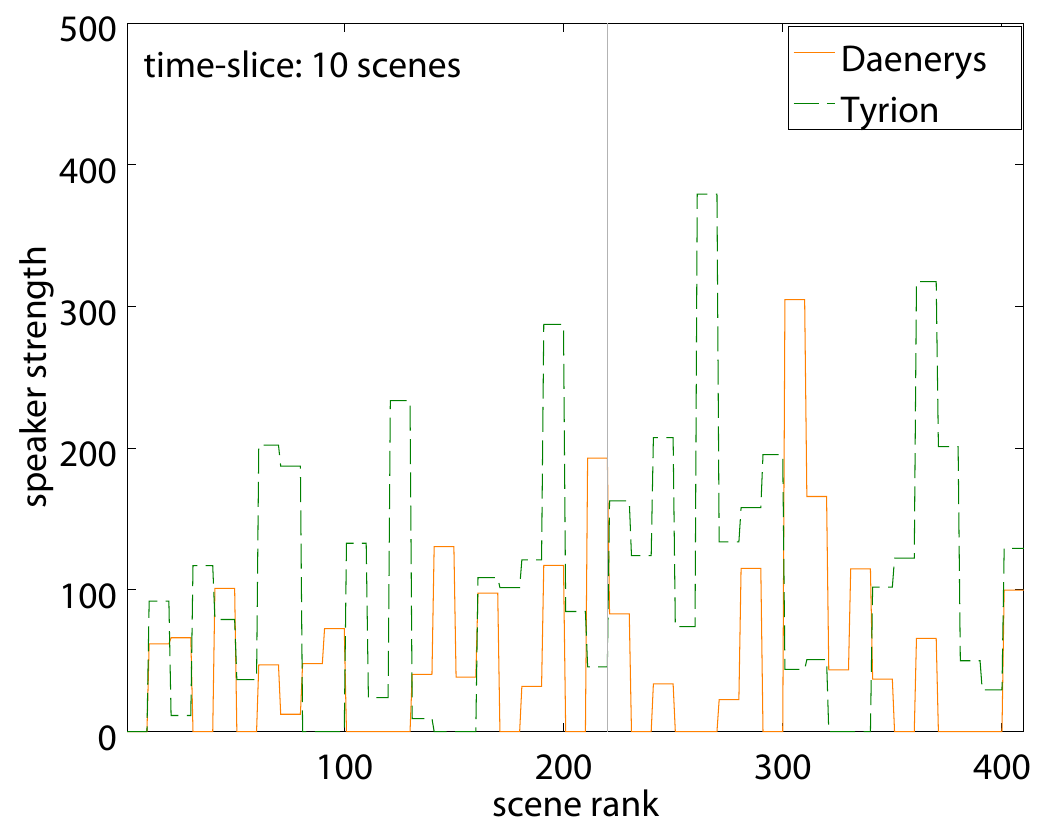}
  }
  \\
  \subfloat[] {
    \includegraphics[width=.43\textwidth]{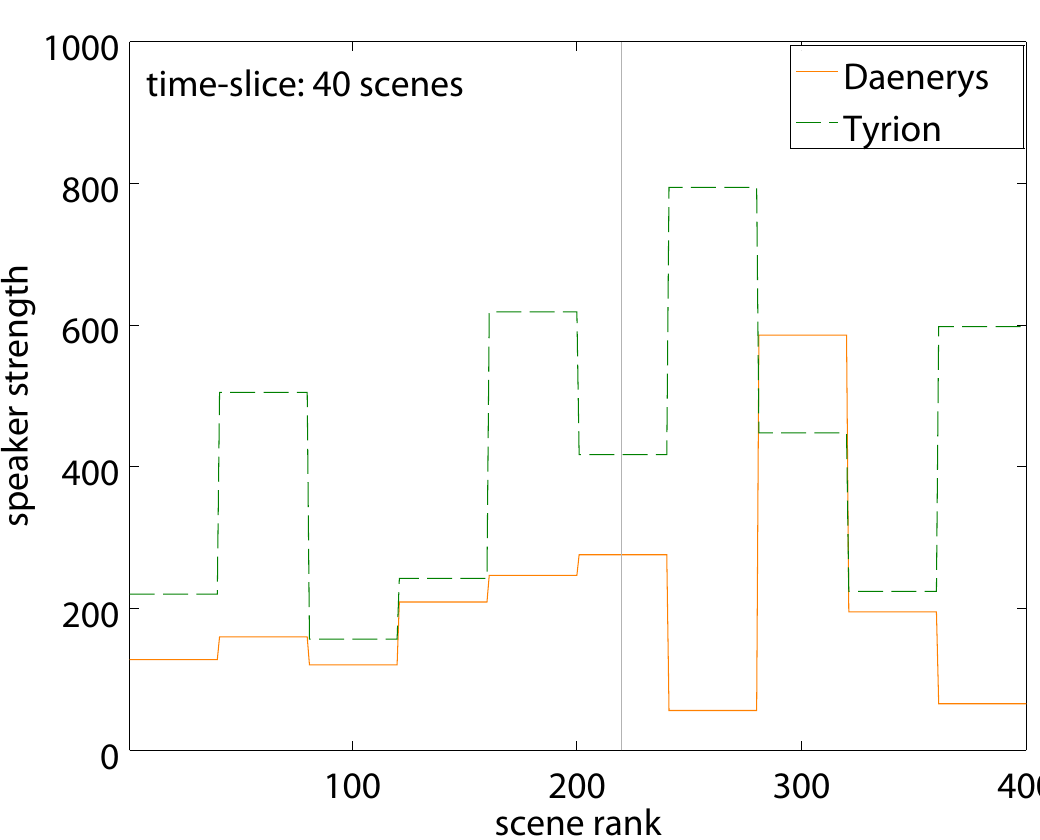}
  }
  \\
  \subfloat[] {
    \includegraphics[width=.43\textwidth]{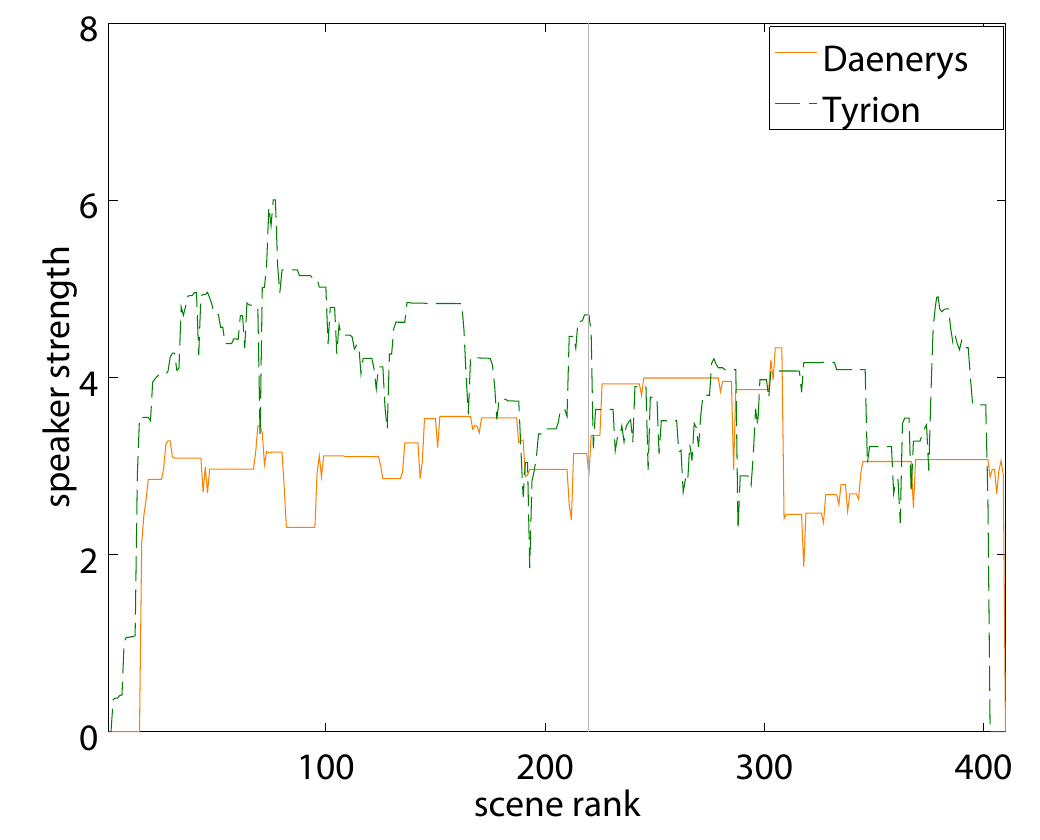}
  }
  \\
  \end{tabular}
  \caption{\label{fig:dyn_strength_got}Strength of two major characters of \textit{Game of Thrones} plotted as a function of the chronologically ordered scenes. From top to bottom: $10$ and $40$ scenes time-slices, and narrative smoothing.}
\end{figure}

We now switch to Daenerys Targaryen and Tyrion Lannister, two major protagonists of \textit{Game of Thrones}. Fig.~\ref{fig:dyn_strength_got} shows how their strengths evolve over the first two seasons of the series, again as a function of the chronologically ordered scenes. The first two plots were obtained through the use of fixed-size observation windows, set to $10$ scenes (around half an episode) for the first and $40$ (about two episodes) for the second. The last plot relies on our narrative-smoothing method. The appearance of Daenerys' storyline onscreen has a relatively slow pace in these seasons (Fig.~\ref{fig:narr_freq}). When the window is too narrow, this creates noisy, irrelevant measurements of her importance (first plot on Fig.~\ref{fig:dyn_strength_got}). It appears very unstable because her storyline alternates with many others on the screen. A wider observation window (second plot of the same figure) is more likely to cover successive occurrences of Daenerys in the narrative, but, unlike our narrative smoothing method, prevents us from locating precisely the scenes responsible for Tyrion's current importance: for instance, a local maximum in Tyrion's strength is reached at scene 220 (third plot on Fig.~\ref{fig:dyn_strength_got}), just after a major narrative event took place~--~the nomination of Tyrion as the King's Counselor (vertical line). Such an event remains unnoticed when accumulating the interactions during too large time-slices (second plot on Fig.~\ref{fig:dyn_strength_got}).

Fig.~\ref{fig:dyn_strength_got} also reveals an important property of our way of building the dynamic network. Because the past (resp. future) occurrences of a particular relationship are still (resp. already) active as long as the involved characters do no interact with others in the meantime, the respective strengths of the main characters of the story appear remarkably balanced. Whereas Tyrion looks much more central than Daenerys in the time-slice based dynamic networks, whatever the size of the observation window, Daenerys is nearly as central as Tyrion in the network based on our narrative smoothing method: few of her acquaintances are shown onscreen as interacting with others, whereas the story, by focusing more frequently on Tyrion, also unveils more extensively his social network, even when active without him (especially after scene 220).

Our results confirm that cumulative networks, by neglecting the temporal dimension, tend to completely miss punctual changes in the importance of certain characters relatively to the plot. The time-slice based methods can handle the network dynamics, however our observations illustrate that they cannot properly tackle the narrative issue we described in Subsection~\ref{subsec:time_slice}. The choice of an appropriate time window, is a particularly sensitive point. By comparison, narrative smoothing captures the state of a relationship at any moment of the plot, using a time scale which directly depends on the narrative pace of the considered series. This allows to finely evaluate the degree of instantaneous involvement of any character in the plot.

%%%%%%%%%%%%%%%%%%%%%%%%%%%%%%%%%%%%%%%
\subsection{The relationships}
We now consider relationships between pairs of characters, instead of single individuals. We characterize each relation depending on its weight, \textit{i.e.} the amount of time the characters talked to each other, either cumulated over time-slices, possibly consisting of the whole set of episodes, or smoothed with respect to the narrative. Like in the previous subsection, we focus on relationships of particular interest.

\begin{figure}[!t]
  \centering
  \begin{tabular}{c}
  \subfloat[] {
    \includegraphics[width=.43\textwidth]{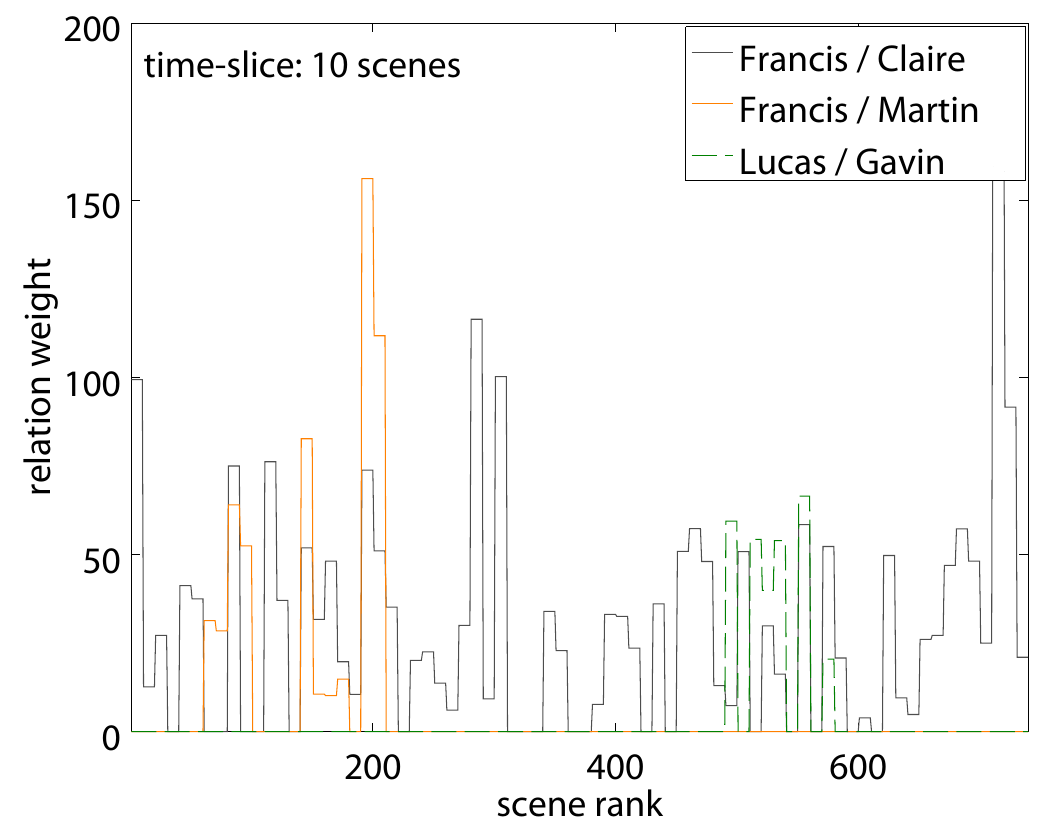}
  }
  \\
  \subfloat[] {
    \includegraphics[width=.43\textwidth]{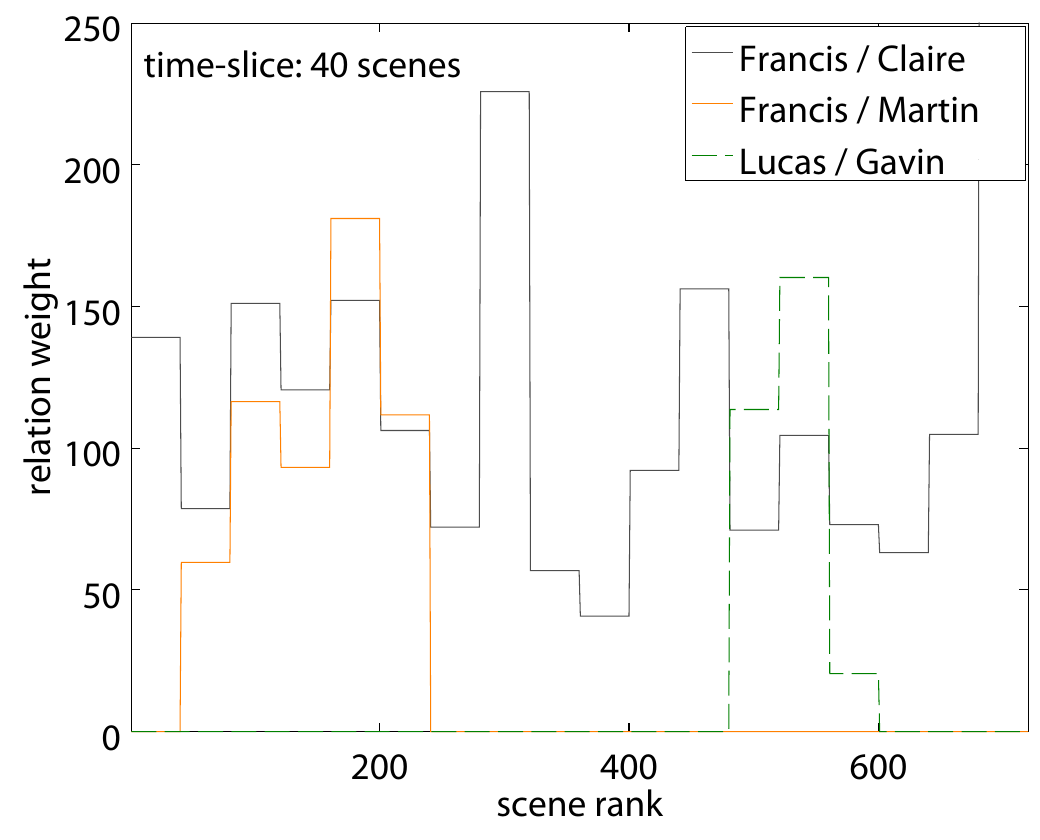}
  }
 \\ 
  \subfloat[] {
    \includegraphics[width=.43\textwidth]{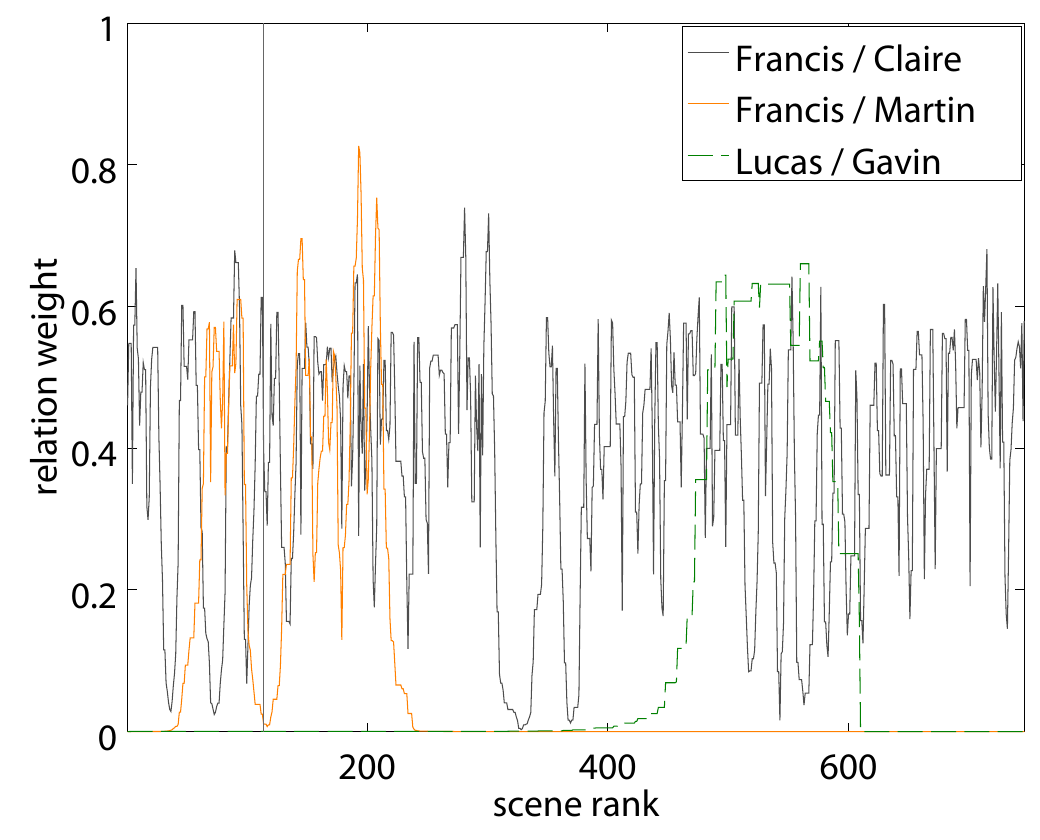}
  }
  \\
  \end{tabular}
  \caption{\label{fig:dyn_weight_hoc}Weight of three relationships between five characters of \textit{House of Cards} plotted as a function of the chronologically ordered scenes. From top to bottom: $10$ and $40$ scenes time-slices, and narrative smoothing.}
\end{figure}

Let us consider two relationships in~\textit{House of Cards}, representative of two sub-stories: the first one corresponds to a narrative sequence in the storyline related to the main character Francis Underwood~--~his fight with a former ally, the unionist Martin Spinella; the second one is a similar subplot, but related to a secondary character, not as frequently present in the narrative, the journalist Lucas Goodwin, who requests the help of the hacker Gavin Orsay to investigate on Francis. Though locally important in these two sub-stories, neither of these relationships lasts long enough to be noticed in the cumulative network, as resulting from the first two seasons of the series (\textit{cf.} Fig.~\ref{fig:HoC_cumul} in the Appendix): the interaction time amounts to 562 seconds for the relation between Francis and Martin, and to 294 seconds for the relation between Gavin and Lucas. These total interaction times remain quite small compared to the central relation between Claire and Francis, amounting to 2,319 seconds.

Nonetheless, once plotted as a function of the chronologically ordered scenes (Fig.~\ref{fig:dyn_weight_hoc}), the respective weights of these relationships in the narrative look quite different, whatever the weighting scheme. Both sub-stories, the one based on the relation between Francis and Martin and the one based on the relation between Lucas and Gavin, turn out to be locally as important as the long-term sub-story based on the relation between the two main characters Claire and Francis. 

Furthermore, all three ways of monitoring these relationships over time are not equivalent: agglomerating the interactions within short time-slices (first plot on Fig.~\ref{fig:dyn_weight_hoc}) makes us miss the continuity of Lucas/Gavin's sub-story, which occurs \textit{in the narrative} at a slower rate than the sub-stories related to Francis. Conversely, large time-slices (second plot on Fig.~\ref{fig:dyn_weight_hoc}) allow to capture this sub-story, but agglomerate the two main stages of the relation Francis/Martin: before becoming an enemy, Martin is first an ally of Francis; these two parts in the relation correspond to well-separated stages in the narrative, that too large time-slices tend to merge, whereas the separation remains clear when using our narrative smoothing method (materialized by a vertical line on the third plot of Fig.~\ref{fig:dyn_weight_hoc}). 

Our results confirm that cumulative network are inappropriate when attempting to capture punctual sub-stories supported by specific relationships. Moreover, though much more appropriate to such a task, the time-slice approaches suffer from a major drawback: once fixed, the time slice cannot adapt to the variable rates at which the sub-stories appear in the narrative. By overcoming the narrative contingencies, our narrative smoothing approach allows to monitor more accurately over time any relationship, whatever the way the narrative focuses on it.

\begin{figure}[!t]
  \centering
  \includegraphics[width=.43\textwidth]{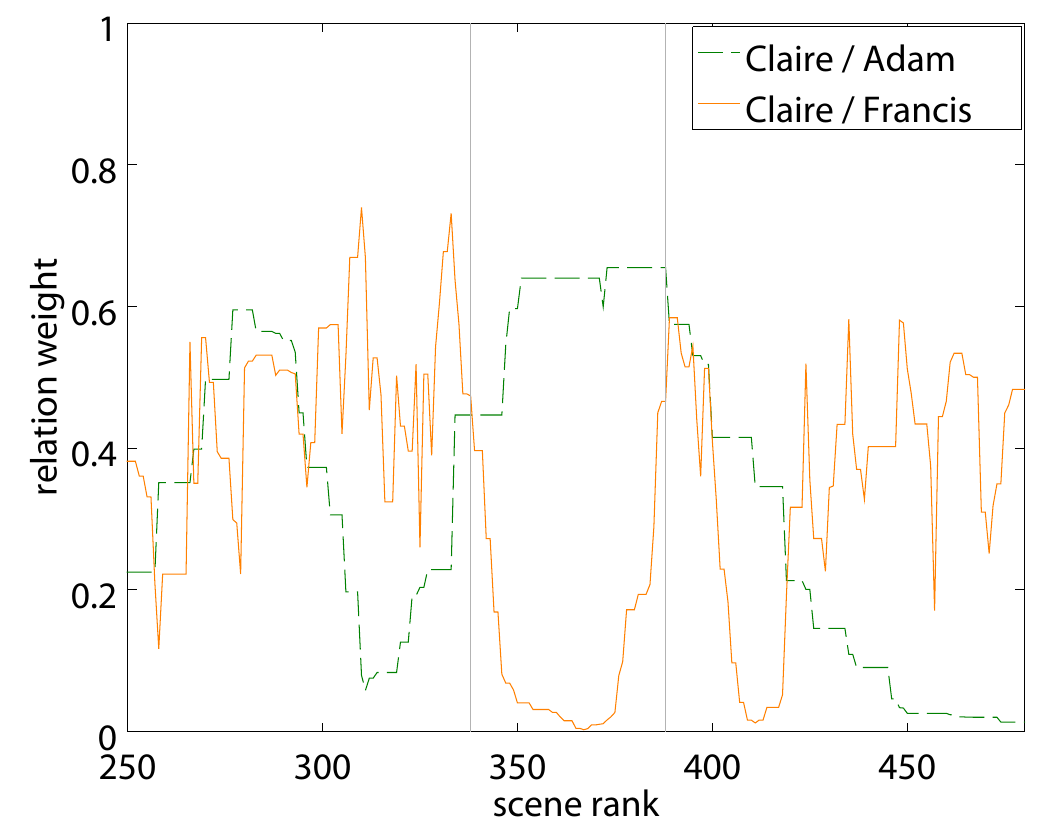}
  \caption{\label{fig:dyn_weight_hoc_bis}Weight of two relationships between three characters of \textit{House of Cards} plotted as a function of the chronologically ordered scenes and based on narrative smoothing.}
\end{figure}

Fig.~\ref{fig:dyn_weight_hoc_bis} illustrates the possible insight our narrative smoothing approach can give on important sub-stories.

As can be seen, two relationships between three characters of \textit{House of Cards} are considered: the relationship between Claire and her husband on the one hand and the relationship between Claire and her former lover Adam Galloway on the other hand. In the cumulative graph (\textit{cf.} Fig. \ref{fig:HoC_cumul} in the Appendix), the interaction time between the two legal spouses is far more important than between Claire and Adam. Nonetheless, once considered over the narrative, it is clear that the relationship between Claire and Adam is locally much stronger than with her husband: between the scenes 338 and 388 (vertical lines on the plot), both relationships are complementary and suggest a specific sub-plot where Claire is much closer to Adam than to Francis.

\begin{figure}[!t]
  \centering
  \includegraphics[width=.43\textwidth]{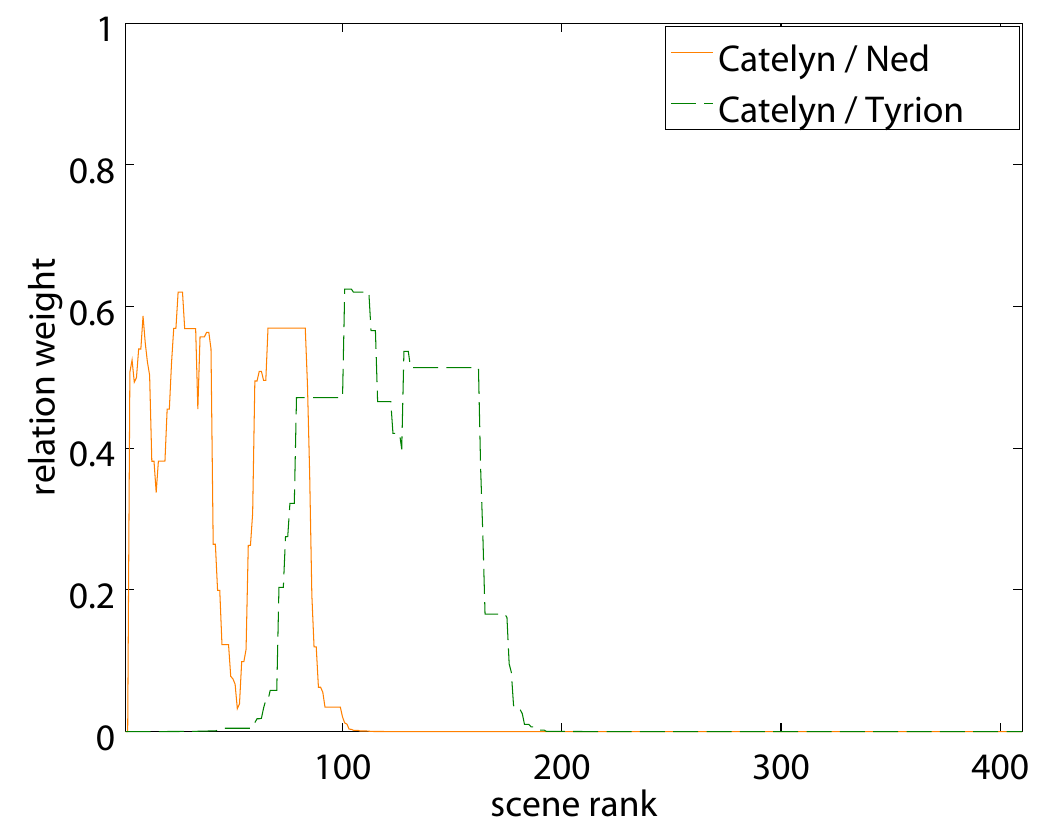}
  \caption{\label{fig:dyn_weight_got_bis}Weight of two relationships between three characters of \textit{Game of Thrones} plotted as a function of the chronologically ordered scenes and based on narrative smoothing.}
\end{figure}

Let us go go back once again to \textit{Game of Thrones} and its complex plot. Fig.~\ref{fig:dyn_weight_got_bis} focuses on two relationships between three characters: Catelyn Stark and Ned Stark on the one hand, Catelyn Stark and Tyrion Lannister on the other. 

Neither of these relationships would be considered as a major one from the cumulative graph at the end of the first two seasons (\textit{cf.} Fig.~\ref{fig:GoT_cumul} in the Appendix). Nevertheless, once dynamically considered, they both correspond to two successive sub-stories in the first season of \textit{Game of Thrones}. As can be seen, our narrative smoothing approach even allows to separate two steps in the relationship between Catelyn and Ned: the first step of their relationships takes place in Winterfell; Ned then leaves Catelyn there and goes on his own to Kings Landing, freshly named as the King's Counselor (around the scene 65), before Catelyn joins him there. Catelyn and Tyrion start interacting with each other after Catelyn leaves Kings Landing to Winterfell and is well preserved once monitored according to our method, though shown in the narrative at a quite slow and irregular pace.

%%%%%%%%%%%%%%%%%%%%%%%%%%%%%%%%%%%%%%%%%%%%%%%%%%%%%%%%%%%%%%%%%%%%%%%%
\section{Conclusion and Perspectives}
\label{sec:conclu}
In this paper, we described a novel way of monitoring over time the state of the relationships between characters involved in the usually complex plots of modern \textsc{tv} series. The two methods previously used for this purpose are the cumulative approach, consisting in integrating every relation over the whole considered period of time, and the time-slice approach, consisting in breaking down the time-line into smaller discrete chunks. The first one turns out to be relatively inefficient for investigating complex storylines and a dynamic perspective is more appropriate. The second one complies with this constraint, but defining an appropriate size for the observation window is a very difficult task and constitutes a major drawback: the plots of modern \textsc{tv} series usually consist in parallel storylines shown sequentially onscreen at an unpredictable frequency. As a main consequence, the narrative disappearance in the current scene of some past relationship can usually not be interpreted as a real disappearance, which invalidates the time-slice approach. To address this issue, we chose to smooth the narrative sequentiality, by considering that the relation between interacting speakers remains active as long as neither of them speaks with others; if so, such separate interactions result in a progressive dissolution of the past link. Symmetrically, the imminence of the next occurrence of the relationship has to strengthen the link. We experimentally compared our method, which we call \textit{narrative smoothing}, to both mentioned approaches on three recent popular \textsc{tv} series. Though exploratory and qualitative, our results show that our method leads to more relevant results than both other methods, when it comes to instantaneously monitoring the importance of a particular character or of a specific relationship at some point of the story.

The way some characters temporarily aggregate at some point of the story in a community-like structure suggests some narrative sequences result in the stabilization, possibly temporarily, of certain areas in the network. By automatically detecting such a narrative stabilization of some groups of relationships, it should be possible to split the whole story into sub-stories, without assuming a static, predefined, community structure. Finally, the statistical properties of such a dynamic network, as based on the smoothing of the narrative, have still to be studied: the relative balance between the important characters suggests, for instance, that the traditional heavy-tailed degree distribution may not stand in this case.

\newpage

\section*{Acknowledgments}

This work was supported by the French National Research Agency
(\textsc{anr}) \textsc{gafes} project (\textsc{anr}-14-CE24-0022) and
the Research Federation \textit{Agorantic}, University of Avignon.



\bibliographystyle{IEEEtran}
\bibliography{IEEEabrv,base}


% \end{thebibliography}

%%%%%%%%%%%%%%%%%%%%%%%%%%%%%%%%%%%%%%%%%%%%%%%%%%%%%%%%%%%%%%%%%%%%%%%%
\section{Appendix}
\begin{figure*}[!t]
    \centering
	\includegraphics[width=\textwidth]{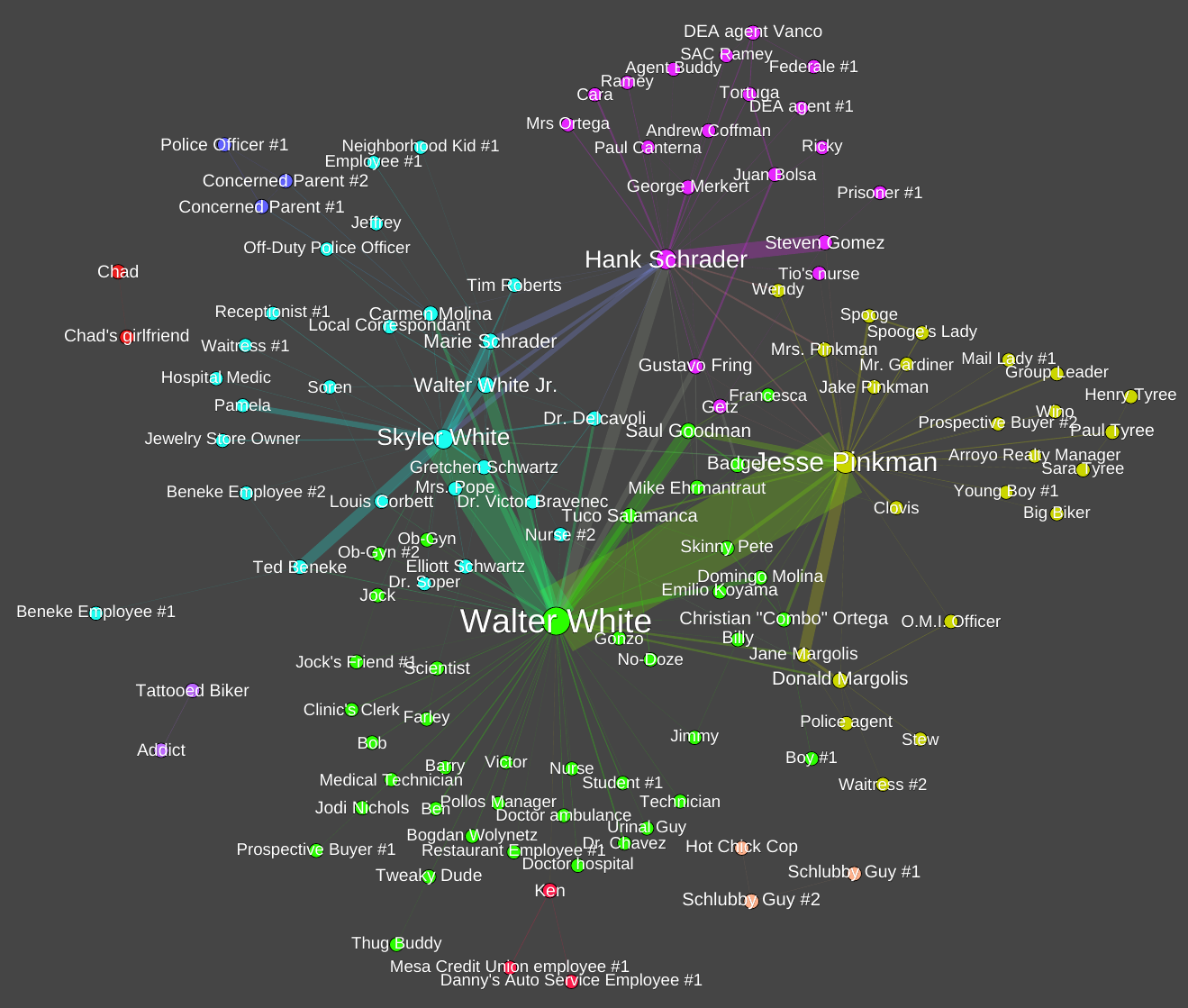}
    \caption{\label{fig:BB_cumul}Cumulative network for the first two seasons of \textit{Breaking Bad}.}
\end{figure*}

\begin{figure*}[!t]
    \centering
	\includegraphics[width=\textwidth]{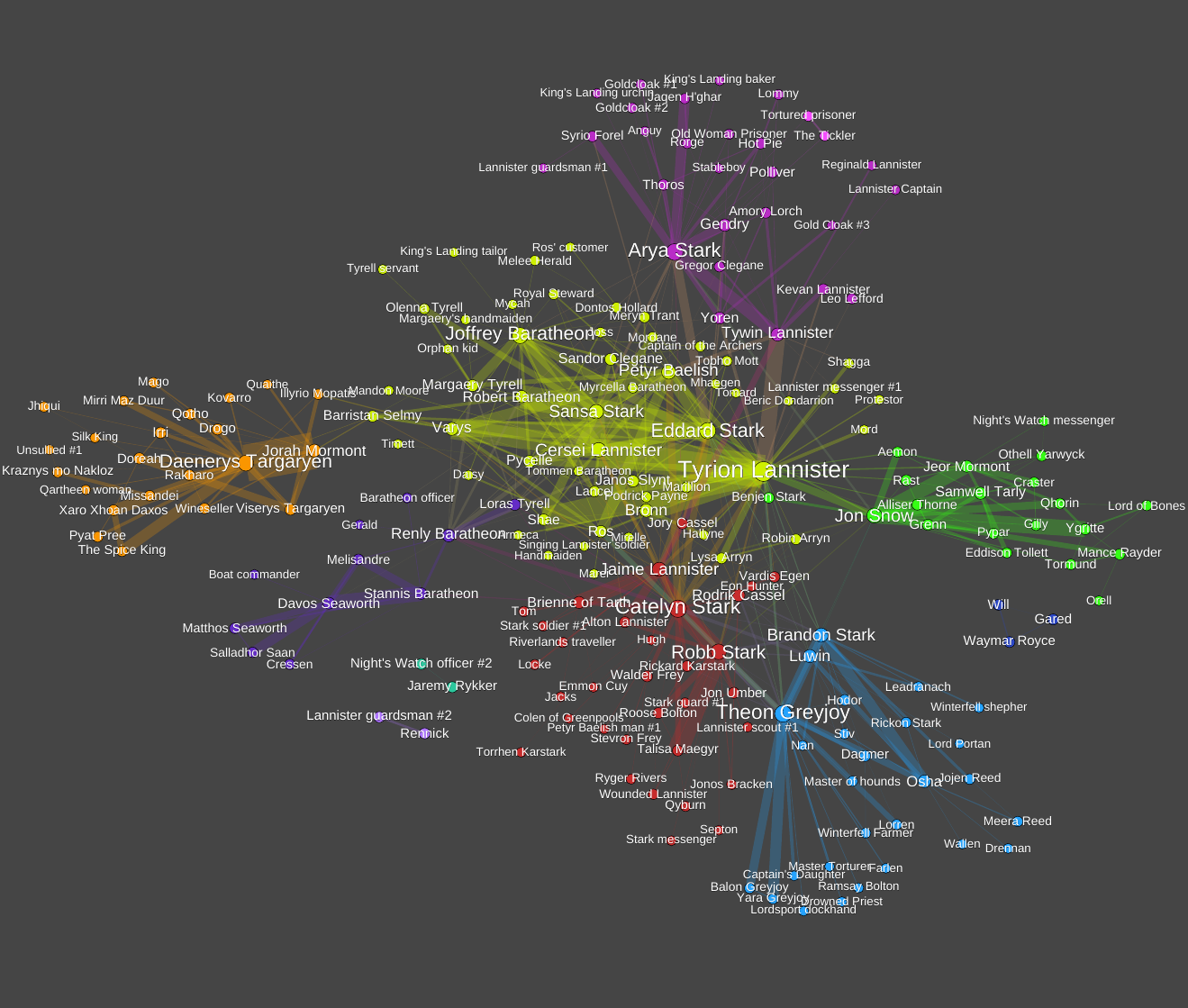}
    \caption{\label{fig:GoT_cumul}Cumulative network for the first two seasons of \textit{Game of Thrones}.}
\end{figure*}

\begin{figure*}[!t]
    \centering
	\includegraphics[width=\textwidth]{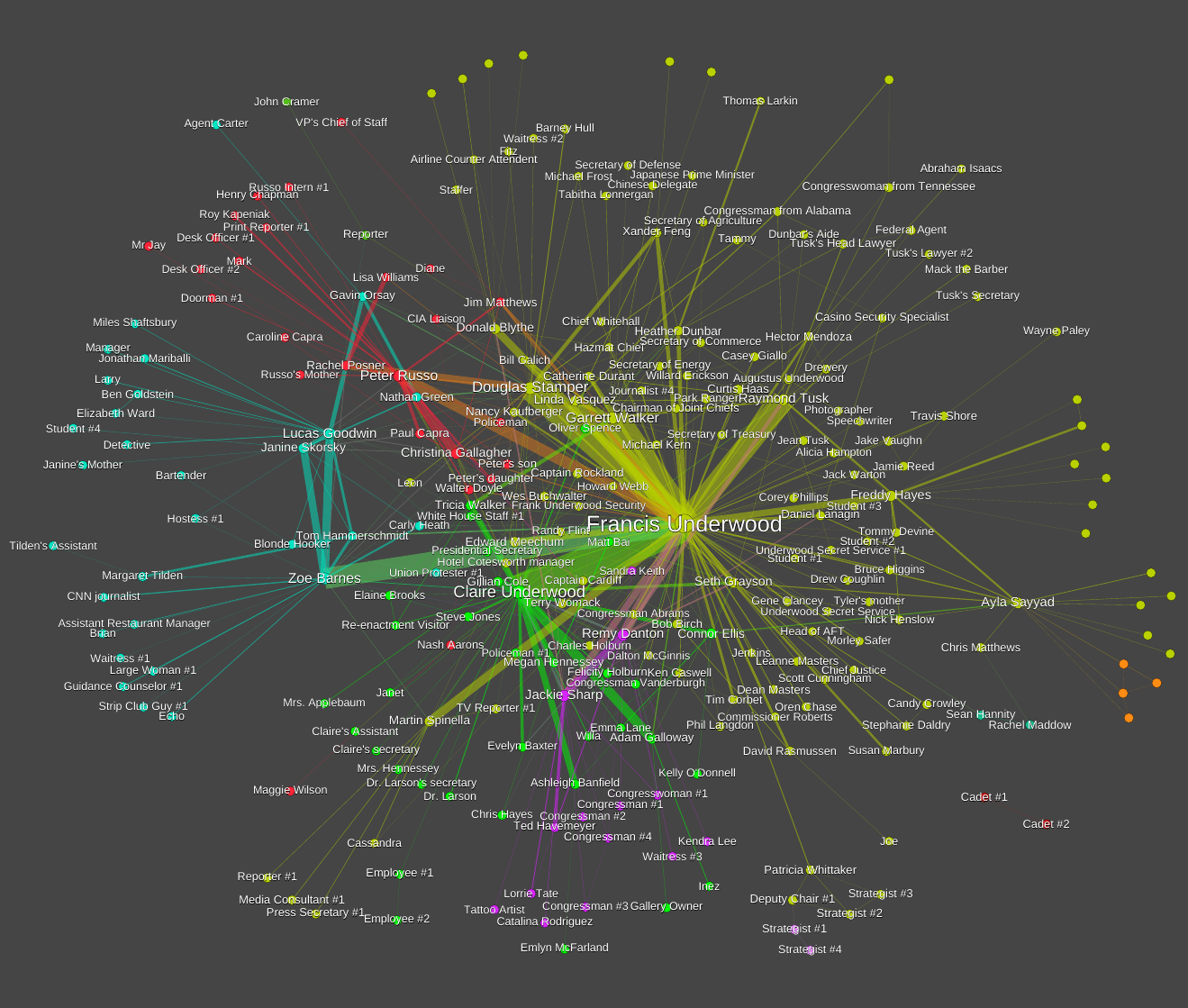}
    \caption{\label{fig:HoC_cumul}Cumulative network for the first two seasons of \textit{House of Cards}.}
\end{figure*}

Figures \ref{fig:BB_cumul}, \ref{fig:GoT_cumul} and \ref{fig:HoC_cumul} show the cumulative networks of Breaking Bad, Game of Thrones and House of Cards, respectively, extracted over their first two seasons.

\end{document}